\documentclass[11pt]{article}
\pdfoutput=1 
\usepackage{jheppub,epsfig,amssymb,amsmath}

\def\bea{\begin{eqnarray}}
\def\eea{\end{eqnarray}}
\def\beq{\begin{equation}}
\def\eeq{\end{equation}}

\preprint{PNUTP-17-A12, APCTP Pre2017-016}

\title{Peccei-Quinn Relaxion}

\author[a]{Kwang Sik Jeong,}
\emailAdd{ksjeong@pusan.ac.kr}
\affiliation[a]{Department of Physics, Pusan National University, Busan 46241, Korea} 
 
\author[b,c,d]{Chang Sub Shin}
\emailAdd{csshin@ibs.re.kr}
\affiliation[b]{Asia Pacific Center for Theoretical Physics, Pohang 37673, Korea}
\affiliation[c]{Department of Physics, Postech, Pohang 37673, Korea}
\affiliation[d]{ Center for Theoretical Physics of the Universe, Institute for Basic Science (IBS), \\Daejeon 34051, Korea}

\abstract{ 
The relaxation mechanism, which solves the electroweak hierarchy problem
without relying on TeV scale new physics, crucially depends on how a Higgs-dependent 
back-reaction potential is generated. 
In this paper, we suggest a new scenario in which the scalar potential induced by the QCD 
anomaly is responsible both for the relaxation mechanism and the Peccei-Quinn mechanism 
to solve the strong CP problem. 
The key idea is to introduce the relaxion and the QCD axion whose cosmic evolutions become
quite different depending on an inflaton-dependent scalar potential. 
Our scheme raises the cutoff scale of the Higgs mass up to $10^7$~GeV, 
and allows reheating temperature higher than the electroweak scale as would be required
for viable cosmology.
In addition, the QCD axion can account for the observed dark matter of the universe as produced
by the conventional misalignment mechanism.  
We also consider the possibility that the couplings of the Standard Model depend on the inflaton
and become stronger during inflation.
In this case, the relaxation can be implemented with a sub-Planckian field excursion of the relaxion
for a cutoff scale below $10$~TeV. 
}
 
\begin{document}
\maketitle
   
\section{Introduction}

As sensitive to unknown high energy physics, the mass of the Higgs boson in the Standard Model (SM) 
seems unnatural and would require an explanation unless new physics beyond the SM appears around 
TeV scale. 
The LHC null results would thus indicate that we need a new approach to the electroweak hierarchy problem. 
In this context, it has recently been proposed to consider 
dynamical evolution during early universe that drives the Higgs mass to a value much smaller 
than the cutoff scale of  the theory~\cite{Graham:2015cka}.
This relaxation mechanism is based on the interplay between the Higgs field $h$ and an axion-like 
scalar $\phi$,
arising from cosmological evolution such that $\phi$ slowly rolls during inflation while scanning 
the effective Higgs mass-squared term over a large range 
until it meets barriers formed by electroweak symmetry breaking.
 
The relaxion mechanism crucially relies on how to generate Higgs-dependent barriers for the relaxion
$\phi$, which are to stop the relaxion from rolling and set the Higgs mass to a naturally small value. 
A natural source is the QCD anomaly, for which however the model is generally subject to severe constraints 
coming from the experimental bound on the strong CP phase. 
Alternatively one can consider a non-QCD source of barriers.
In such case, the gauge invariance requires that a barrier potential be proportional to $h^2$,
implying that it should be generated at a scale not much above the electroweak scale since otherwise
closed Higgs loops would produce Higgs-independent high barriers and spoil the relaxation.
The difficulties in each case can be resolved if the relaxion sector is extended to include more scalars
or to be coupled to the inflation sector.

In this paper we present a new possibility where the QCD anomaly is responsible both for
the relaxation mechanism to explain the smallness of the Higgs mass and for the Peccei-Quinn (PQ) mechanism 
to solve the strong CP problem~\cite{ Peccei:1977hh,Kim:2008hd}. 
The idea is that both the relaxion and the QCD axion couple to the QCD anomaly, and in addition each of them
couples to a hidden confining gauge anomaly.
A scalar potential induced by hidden gauge anomalies changes its form due to the roll of an inflation field
in such a way that it effectively depends only on the QCD axion during inflation but only on the relaxion after inflation.
This allows the QCD-induced potential to serve as Higgs-dependent barriers for the relaxion during inflation 
but as a potential to fix the QCD axion after inflation.
As a result, the relaxation mechanism is well implemented and can raise the cutoff scale of the Higgs boson 
up to about $10^7$\,GeV.

It is interesting to note that in our scheme the relaxion has a negligible mixing with the Higgs boson, and 
can obtain a heavy mass compared to other scenarios because it is stabilized by a hidden confining 
force after inflation regardless of the Higgs mass.
This indicates that reheating temperature higher than the electroweak scale is compatible with
the relaxation as long as the hidden confining scale is high enough, which would be important for
viable cosmology.
We also note that the QCD axion does not participate in selecting the Higgs mass but is still important
since it dynamically cancels the strong CP phase.
In addition, the QCD axion can account for the observed dark matter of the universe. 

The relaxation mechanism cosmologically sets the Higgs mass to a small value, but generally at the price 
of a huge excursion of the relaxion and a long period of inflation.
It has been noticed that the clockwork mechanism provides a technically natural framework to arrange 
a long excursion of the relaxion via a collective rotation of multiple axions~\cite{Choi:2015fiu,Kaplan:2015fuy}.
On the other hand, the duration and scale of inflation are constrained essentially by the height of
relaxion barriers during inflation.
In our scheme the constraints on inflation can thus be alleviated if SM couplings become stronger during 
inflation so that the QCD-induced potential is enhanced.
Then it turns out that the Higgs mass selected by the relaxion can be kept after inflation for a cutoff scale 
lower than about $10^4$\,GeV.  
 
The outline of this paper is as follows. 
In Section 2 we review the relevant features of the dynamical relaxation mechanism generating
a naturally small electroweak scale.
In Section 3 we construct a simple model where the relaxion and the QCD axion play their respective
roles via the QCD anomaly, and examine in detail how the relaxation works.
The constraints on inflation are examined in Section 4 for the case where SM couplings become stronger 
during inflation.
Conclusions are given in Section 5.

\section{Cosmological Relaxation of the Electroweak Scale}

In this section we briefly review the cosmological relaxation of the electroweak scale. 
The relaxation mechanism generating a naturally small electroweak scale is implemented by the relaxion 
$\phi$ that has a scalar potential
\bea
V = V_0(\phi)  + m^2_h(\phi) h^2 + V_{\rm br} (\phi,h),
\eea
where $h$ is the Higgs field. 
The sliding potential $V_0$ makes the relaxion slowly roll down during inflation while scanning the effective 
Higgs mass-squared term $m^2_h$ over a large range.
The last term $V_{\rm br}$ appears after electroweak symmetry breaking and provides barriers stopping
the evolution of the relaxion. 
The three potential terms take the form
\bea
V_0(\phi) &=& M^4 \left(
-c_1 \frac{\phi}{F} + c_2 \frac{\phi^2}{F^2} +  \cdots \right), 
\nonumber \\
m^2_h(\phi) &=& 
M^2 \left(	k_0 - k_1 \frac{\phi}{F}  + \cdots \right), 
\nonumber \\
V_{\rm br}(\phi,h) &=& -\mu^4_{\rm br}(h) \cos\left(\frac{\phi}{f}\right),
\eea
for positive coefficients $c_i$ and $k_i$ of order unity.
Here $M$ is the cutoff scale of the theory, and $M/F \ll 1$ parameterizes
the breaking of shift symmetry $\phi \to \phi + 2\pi f$.
Note that the parameter $c_1$ is bounded from below
\bea
c_1 \gtrsim \frac{k_1}{16\pi^2},
\eea
in the presence of the $k_1$-term, because the scalar potential generally receives a contribution 
from closed Higgs loops.

For the relaxation mechanism to work, the inflationary energy density should be larger than the 
change of the energy density in the relaxion sector, $V_0 < H^2_i M^2_{Pl}$, and the relaxion 
evolution is dominated by classical rolling,
$\dot \phi/H_i \approx  \partial_\phi V_0/H^2_i > H_i$, implying
\bea
\label{condition1}
\frac{\sqrt{V_0}}{M_{Pl}} < H_i< (\partial_\phi V_0)^{1/3},
\eea
where $H_i$ is the Hubble scale during inflation.
In addition, inflation should last long enough for the relaxion to scan the Higgs mass-squared term 
from positive to negative, which generally requires a large number of $e$-folds 
\bea
N_e  \gtrsim \frac{H^2_i}{\partial_\phi V_0} F ,
\eea
for $k_i$ of order unity,
because the relaxion changes by an amount $\Delta \phi \sim (\dot \phi/H_i) N_e$ during inflation.
%
To stop the relaxion from rolling, high enough barriers for $\phi$ should be formed during
inflation, implying that one needs
\bea
\label{condition2}
\partial_\phi V_0 \sim \partial_\phi V_{\rm br},
\eea
at the time when the Higgs vacuum expectation value is near its SM value.
The QCD axion solving the strong CP problem can play the role of the relaxion, for which case
the barrier potential is induced by the QCD anomaly and has 
\bea
\mbox{QCD :} &&  
 \mu_{\rm br}^4(h)=  y_u \Lambda_{\rm QCD}^3 h,
\eea
where $y_u$ is the up quark Yukawa coupling, and $\Lambda_{\rm QCD}\sim 0.1$\,GeV is 
the QCD scale.  
However, the potential $V_0$ required for slow rolling of the relaxion during inflation
produces too large strong CP violation. 
This problem can be avoided if the slope of $V_0$ arises from a coupling to the inflaton and 
dynamically decreases after inflation~\cite{Graham:2015cka}, or conversely if  
the slope of $V_{\rm br}$  increases after inflation~\cite{Nelson:2017cfv}.
Another way is to consider a non-QCD model where barriers are produced by a hidden strong  
gauge interaction. 
In such case, one has
\bea
\mbox{non-QCD :} &&  
\mu_{\rm br}^4(h) = \Lambda_{\rm hid}^2 h^2,
\eea
where $\Lambda_{\rm hid}$ is the confining scale of the hidden gauge interaction,
and $\mu^4_{\rm br} \propto h^2 \in H^\dagger H$ reflects that the new sector should couple to 
the Higgs doublet $H$ in a gauge-invariant way.  
The strong CP phase is then cancelled by the QCD axion independently of the relaxation mechanism.
However, the barrier potential depends on the Higgs field as $\mu^4_{\rm br} \propto h^2$, and so suffers from 
the coincidence problem, that is, a requirement that the hidden confining scale should be around 
the electroweak scale since otherwise closing Higgs loops would induce large barriers even before 
electroweak symmetry breaking and spoil the relaxation mechanism.
This problem may be avoided in an extended model with multiple axions where $h$-independent barriers 
for the relaxion are suppressed by the double-scanning mechanism working under certain assumptions
on the involved phases~\cite{Espinosa:2015eda}.

\section{Peccei-Quinn Relaxion}

The viable region of parameter space for the relaxion mechanism crucially depends on the origin
of barriers for the relaxion.
Here we present a model involving two axions, the relaxion ($\phi$) and the QCD axion ($a$), 
where the QCD-induced potential plays an essential role in the relaxation as well as 
in the PQ mechanism. 
In our scenario, both $\phi$ and $a$ couple to the QCD anomaly, but nonetheless play their 
respective roles in cosmologically relaxing the electroweak scale and dynamically 
cancelling the strong CP phase. 
This is achieved via the barrier potential,
\bea
V_{\rm br}(a,\phi,h) = -\mu_{\rm br}^4(h)
\cos\left(\frac{a}{f_a} + \frac{\phi}{f} \right)
+ \Delta V_{\rm br}(a,\phi),
\eea  
where the first term comes from the QCD anomaly,
\bea
\mu^4_{\rm br}(h)= y_u \Lambda^3_{\rm QCD} h
\simeq (0.1{\rm GeV})^4(h/v),
\eea 
with $v=246$~GeV being the vacuum expectation value of $h$ in the present universe.
The barrier potential includes an additional term, $\Delta V_{\rm br}$, which changes its form during 
and after inflation due to the roll of the inflaton: 
\bea
\label{deltaV}
\Delta V_{\rm br} = 
\mu^4_a(\sigma) 
\cos\left(n_a \frac{a}{f_a} \right) 
+ \mu^4_\phi(\sigma) \cos\left( n_\phi \frac{\phi}{f} \right),
\eea 
in which $\mu_a$ and $\mu_\phi$ are a function of the inflaton field $\sigma$, and both evolve with time 
but in the opposite way
\bea\label{eq:br_condition}
\mu_a^4(\sigma=\sigma_0) \ll  &\mu^4_{\rm br}(h=v)& \ll \mu^4_a(\sigma= \sigma_{\rm inf}),
\nonumber \\ 
\mu_\phi^4(\sigma=\sigma_{\rm inf} ) \ll  &\mu^4_{\rm br}(h=v)& \ll \mu^4_\phi (\sigma = \sigma_0),
\eea 
under the assumption that the inflaton slowly rolls with a large field value $\sigma_{\rm inf} \gg M$ 
during inflation, and is stabilized at the true minimum  $\sigma = \sigma_0 \ll M$ after inflation.   
The rational constants $n_a$ and $n_\phi$ will be set to be unity hereafter for simplicity as our results 
do not depend much on them.  
Note that the change of the energy density in the barrier sector should be much 
smaller than $H^2_i M^2_{Pl}$ in order not to affect the inflation dynamics, which puts an upper bound
on $\mu_a(\sigma=\sigma_{\rm inf})$ and $\mu_\phi(\sigma=\sigma_0)$.  
 
Let us discuss how to generate a barrier potential relying on the evolution of the inflation field
as required for our scenario to work. 
As the case of the Higgs-dependent axion potential, a $\sigma$-dependent back-reaction potential for $a$
and $\phi$ can be generated non-perturbatively.  
We consider a hidden confining group $G_a\times G_\phi$, where the shift symmetries 
\bea
{\rm U}(1)_a &:& a\to a + {\rm constant},
\nonumber \\
{\rm U}(1)_\phi &:& \phi\to \phi + {\rm constant},
\eea
are non-perturbatively broken by the $G_a$ and $G_\phi$ anomaly, respectively, in the presence
of hidden vector-like quarks $Q_a+ Q_a^c$ charged under U$(1)_a$ and $G_a$, and
$Q_\phi+Q^c_\phi$ charged under U$(1)_\phi$ and $G_\phi$. 
The hidden quarks have U$(1)_i$-preserving Yukawa interactions 
\bea
\sum_i y_i S_i Q_i Q^c_i,
\eea
for $i=a,\phi$.
The scalar potential $\Delta V_{\rm br}$ can be obtained if $S_a$ and $S_\phi$ undergo non-trivial time evolution 
due to a coupling to the inflaton. 
A simple way is to consider an inflaton-dependent mass-squared term whose sign is flipped during and after inflation:
\bea
\label{inflaton-dependent-dynamics}
V = (M^2 - \kappa_a \sigma^2) |S_a|^2
-(M^2 - \kappa_\phi \sigma^2) |S_\phi|^2
+ |S_a|^4 + |S_\phi|^4 + \Delta V(S_i,e^{ia/f_a},e^{i\phi/f}), 
\eea
with positive constants $\kappa_i$ lying in the range
\bea
\label{constraint-kappa}
\left(\frac{M}{\sigma_{\rm inf}}\right)^2 < \kappa_ i \ll 1,
\eea
for which $S_i$ are fixed at quite different values during and after inflation but without disturbing
the inflaton dynamics. 
In the scalar potential, we have omitted other constant coefficients of order unity for simplicity,
and $\Delta V$ includes U$(1)_i$-preserving interactions of $S_i$ to $a$ and $\phi$ responsible for
heavy masses of the U$(1)_a$-invariant combination of $\arg(S_a)$ and $a$, and 
the U$(1)_\phi$-invariant combination of $\arg(S_\phi)$ and $\phi$.  
Note that the compositions of the QCD axion and the relaxion change,
$a + \frac{\langle S_a \rangle^2}{f_a} \arg(S_a) \to a$ and 
$\phi \to \phi + \frac{\langle S_\phi \rangle^2}{f} \arg(S_\phi) $, neglecting order unity coefficients. 
Owing to the inflaton-dependent scalar masses, the hidden quarks obtain masses according to
\bea
m_{Q_a}(\sigma_{\rm inf}) &=& y_a \sqrt{ \kappa_a \sigma^2_{\rm inf} - M^2 },
\nonumber \\
m_{Q_\phi}(\sigma_{\rm inf}) &\simeq& 0, 
\eea
during inflation, whereas they have
\bea
m_{Q_a}(\sigma_0) &\simeq& 0,
\nonumber \\
m_{Q_\phi}(\sigma_0) &=& y_\phi \sqrt{ \kappa_\phi \sigma^2_{\rm inf} - M^2 },, 
\eea
after inflation. 
The hidden gauge anomalies to which the relaxion and the QCD axion couple generate 
an inflaton-dependent back-reaction potential of the form, Eq~(\ref{deltaV}), with the overall
size determined by
\beq
\mu_i^4(\sigma)= {\rm Min}\Big[m_{Q_i}(\sigma), \Lambda_i \Big]\, \Lambda_i^3,
\eeq
for $i=a,\phi$, where $\Lambda_i$ denotes the confining scale of the corresponding hidden gauge group.
Therefore the potential $\Delta V_{\rm br}$ can have the required properties, Eq~(\ref{eq:br_condition}),
if the hidden confining scales are higher than $\Lambda_{\rm QCD}$, under the assumption that
U$(1)_a$ and U$(1)_\phi$ are spontaneously broken at a scale much higher
than $M$ by scalar fields other than $S_i$.
Note that the hidden gauge groups confine at different scales during and after inflation:
\bea
\Lambda_a(\sigma_{\rm inf}) &>& \Lambda_a(\sigma_0),
\nonumber \\
\Lambda_\phi(\sigma_{\rm inf}) &<& \Lambda_\phi(\sigma_0),
\eea
because the gauge coupling of $G_i$ runs faster (slower) at low energy scales if $Q_i+Q^c_i$ 
become heavier (lighter).   
As discussed below, the above behavior of the hidden sector makes our scheme more natural.

To solve the strong CP problem, the QCD axion should be stabilized mainly by the scalar potential
induced by the QCD anomaly after inflation.
This puts a constraint on the hidden sector if $\Delta V_{\rm br}$ is non-perturbatively generated as 
discussed above, because the $G_a$-anomaly induces a small tadpole term for $S_a$, fixing it 
at $S_a \sim y_a \Lambda^3_a(\sigma_0)/M^2$ after inflation.
As a result, the vacuum of the QCD axion potential is slightly shifted from the origin while generating a strong 
CP phase
\bea
\bar\theta 
&\simeq& \frac{y_a \Lambda^3_a(\sigma_0) \langle S_a \rangle}{\mu^4_{\rm br}(v)}
\sim 10^{-12} 
\left(\frac{ y^{1/3}_a \Lambda_a(\sigma_0)}{0.1\,{\rm GeV}} \right)^6
\left(\frac{M}{10^5\,{\rm GeV}}\right)^{-2},
\eea
which should be smaller than about $10^{-10}$ to avoid the experimental bound.
Here one should note that the $G_a$ gauge force is stronger during inflation than in the present universe
due to the hidden quarks with masses $\propto \langle S_a \rangle$. 
For $\Lambda_a(\sigma_0) \ll \Lambda_a(\sigma_{\rm inf})$, it becomes easier to
satisfy the conditions, $\mu^4_{\rm br}(v) \ll \mu^4_a(\sigma_{\rm inf})$ 
and $\bar\theta < 10^{-10}$.

A hidden sector with quarks having inflaton-dependent masses provides a simple framework to 
generate $\Delta V_{\rm br}$.
Alternatively one may consider explicit breakdown of U$(1)_a$ by higher dimensional operators 
that are turned on once the inflaton develops a vacuum expectation value.
The dependence of $\Delta V_{\rm br}$ on the QCD axion can then be explained for instance by  
\bea
\mu^4_a = \epsilon f^3_a \sigma,
\eea
with $\epsilon\ll 1$, because the explicit symmetry breaking effects disappear in the present universe 
if the inflaton drops to zero after inflation.
The scalar potential $\Delta V_{\rm br}$ relies on the relaxion and the QCD axion quite differently in time. 
Another way to get the proper relaxion-dependence is to consider an inflaton-dependent confining scale, 
$\Lambda_\phi = \Lambda_\phi(\sigma)$, which can be realized if the hidden gauge kinetic term
couples to the inflaton. 
What one needs is that the confining scale is larger than $\Lambda_{\rm QCD}$ 
in the present universe, but it is sufficiently low during inflation,
\bea
\Lambda_\phi(\sigma_{\rm inf}) < H_i,
\eea
so that the anomaly-induced relaxion potential is highly suppressed during the inflationary epoch.

Let us now examine how the cosmological evolution of the relaxion chooses an electroweak
scale hierarchically smaller than the cutoff scale of the theory in our scheme.
During inflation, the QCD axion rolls toward the minimum of $\Delta V_{\rm br}$ and 
settles down there after $N_a$ $e$-folds: 
\bea
N_a \sim \frac{H_i\Delta a}{\dot a}
\sim \frac{f_a H^2_i}{\partial_a \Delta V_{\rm br}}.
\eea
For a correct relaxation process, $N_a$ should be much smaller than $ N_e$ so that 
the QCD axion does not affect the relaxion evolution, 
which is the case for  
\bea 
\sqrt{ \frac{f_a}{F} }\,M <  \mu_a(\sigma_{\rm inf}).
\eea
The above is satisfied in the parameter region of our interest basically because 
inflation lasting for a long enough time is required for the relaxion to scan $m^2_h$
from $M^2$ to a small negative value.
For the QCD axion fixed at the minimum of $\Delta V_{\rm br}$, the relaxation 
conditions, Eqs.~(\ref{condition1}) and (\ref{condition2}), lead to
\bea
H_i &<& \left(\frac{\mu_{\rm br}^4(v)}{f}\right)^{1/3} 
\,\simeq\,
0.5\,{\rm MeV}
\left(\frac{f}{10^6\,{\rm GeV}}\right)^{-1/3}
\left(\frac{ \mu_{\rm br}(v) }{0.1\,{\rm GeV}}\right)^{4/3},
\nonumber \\
M &<& \left(\frac{\mu_{\rm br}^4(v)}{f}\right)^{1/6} \sqrt{ M_{Pl} }
\,\simeq\,
3.3\times 10^7{\rm GeV}
\left(\frac{f}{10^6\,{\rm GeV}}\right)^{-1/6}
\left(\frac{ \mu_{\rm br}(v) }{0.1\,{\rm GeV}}\right)^{2/3},
\eea 
with $F$ given by
\bea
\label{Ff-hierarchy}
\frac{F}{f}  &\sim&  \frac{c_1 M^4}{\mu^4_{\rm br}(v)}
\,\simeq\,
6.3\times 10^{21} \left(\frac{c_1}{1/16\pi^2}\right) 
\left(\frac{\mu_{\rm br}(v) }{0.1\,{\rm GeV}}\right)^{-4}
\left( \frac{M}{10^5\,{\rm GeV}}\right)^4.
\eea 
Thus one needs a huge excursion of the relaxion, $F\gg f$, which 
can be realized in a technically natural manner within the clockwork framework.
Combined with the fact that $f$ cannot be much smaller than the cutoff scale  
at which the barrier potential is generated, the above relation leads to 
\bea
M \lesssim 2\times 10^7\,{\rm GeV}.
\eea
One can also find the condition on the number of  $e$-folds, 
\bea
N_e > \left(\frac{F}{M_{Pl}}\right)^2,
\eea
which is necessary for $\phi$ to scan the effective Higgs mass-squared term from large
positive to negative values.
The cutoff scale should not exceed $6\times 10^5$~GeV if one demands, for instance, $N_e <10^{24}$ 
to avoid fine-tuning of the initial condition in the inflation sector.

After inflation, the role of the potential $\Delta V_{\rm br}$ dramatically changes.
For $\mu^4_\phi(\sigma_0) \gg \mu^4_{\rm br}(v)$, the relaxion settles near a minimum of 
$\Delta V_{\rm br}$, and therefore the QCD-induced potential stabilizes $a$, allowing 
it to implement the PQ mechanism solving the strong CP problem.  
One should note that the relaxion vacuum shift due to $\Delta V_{\rm br}$ would change 
the effective Higgs mass-squared term roughly by the amount, $\pm M^2 f/F$, and so it should
be small enough not to spoil the relaxation mechanism.
This requires  
\bea
\frac{F}{f} >  \left(\frac{M}{v}\right)^2.
\eea
The above constraint is satisfied for $F$ given by Eq.~(\ref{Ff-hierarchy}).
The reheating temperature after inflation is also constrained to be lower than the hidden confining scale
\bea
T_{\rm reh} < \Lambda_\phi(\sigma_0),
\eea
since otherwise the relaxion-dependent part in $\Delta V_{\rm br}$ is highly suppressed by thermal effects. 
Because the barriers for the relaxion after inflation are insensitive to the Higgs vacuum expectation value,
reheating temperature higher than the electroweak scale is compatible with the relaxation mechanism 
as long as $\Lambda_\phi$ is sufficiently high.
This would be important for viable cosmology, for instance, as required in many baryogenesis scenarios. 
Let us examine how high $T_{\rm reh}$ can be in our scheme.
The main constraint comes from that a tadpole term induced by the hidden gauge anomaly
slightly displaces $S_\phi$ from the origin, 
$\langle S_\phi \rangle \sim y_\phi \Lambda^3_\phi(\sigma_{\rm inf})/(\kappa_\phi \sigma^2_{\rm inf})$, 
during inflation.
For the relaxation process not to be disturbed, 
one needs $\mu^4_\phi(\sigma_{\rm inf}) \ll \mu^4_{\rm br}(v)$, implying
\bea
\Lambda_\phi(\sigma_{\rm inf}) < 10^3\,{\rm GeV}
\times
\left( \frac{y_\phi}{10^{-2}} \right)^{-1/3}
\left( \frac{\mu_{\rm br}(v)}{0.1\,{\rm GeV}} \right)^{2/3}
\left( \frac{ \sqrt{\kappa_\phi} \sigma_{\rm inf} }{ 10^9\,{\rm GeV} } \right)^{1/3},
\eea 
with $M \lesssim \sqrt{\kappa_\phi} \sigma_{\rm inf}$.
Combined with the fact that $\Lambda_\phi(\sigma_0)$ is larger than $\Lambda_\phi(\sigma_{\rm inf})$, 
the above relation indicates that reheating temperature can be higher than the electroweak scale in 
a large parameter space.

It is worth noting that the QCD axion can account for the dark matter in the universe
because in our scheme it does not participate in the relaxation mechanism for 
$\mu_a^4(\sigma_{\rm inf}) \gg \mu_{\rm br}^4(v)$.
The relic energy density of the QCD axion is determined by the conventional misalignment 
mechanism~\cite{Kawasaki:2013ae,Berkowitz:2015aua,Patrignani:2016xqp}
\bea
\Omega_a h^2 = k_a \left(\frac{f_a}{10^{12}\, {\rm GeV}} \right)^{1.19}
\theta^2_{\rm ini},
\eea
where $k_a = {\cal O}(0.1)-{\cal O}(1)$ for reheating temperature higher than $\Lambda_{\rm QCD}$.
Here $\theta_{\rm ini}$ is the initial misalignment angle of the QCD axion after inflation, 
and it is determined by  $\Delta V_{\rm br}$.
In the present universe, the QCD axion obtains a mass dominantly from the QCD anomaly and is 
fixed at a CP preserving minimum.
On the other hand, the relaxion becomes massive mainly due to the potential induced by
the hidden gauge anomaly, and thus it has different properties from other models~\cite{Choi:2016luu,Flacke:2016szy,Evans:2017bjs}.
First, the relaxion can have a large mass
\bea
m_\phi \simeq \frac{\mu^2_\phi(\sigma_0)}{f}
= 10^2\,{\rm GeV} \left(\frac{\mu_\phi(\sigma_0)}{10^4{\rm GeV}}\right)^2
\left( \frac{f}{10^6\,{\rm GeV}}\right)^{-1},
\eea
for $\mu_\phi(\sigma_0) < M$, and it decays into SM gauge bosons and possibly into hidden sector
particles depending on a model. 
Another crucial difference is that the relaxion, which becomes heavy due to $\Delta V_{\rm br}$, has 
negligibly small mixing with the Higgs boson because the QCD-induced potential 
$\propto h \cos(a/f_a + \phi/f)$ is responsible both for relaxion-Higgs mixing and 
the stabilization of the QCD axion. 
For $f\gg v$, therefore, it would be difficult to detect the relaxion at collider experiments.  

Let us examine a cosmological constraint arising due to tunneling from 
a local minimum $\phi=\phi_\ast$ to the other minima $\phi\simeq \phi_\ast + 2\pi n f$
for an integer $n$.
The slow-rolling potential lifts the vacuum degeneracy of the barrier potential, 
$V_{\rm br} = \mu_{\phi}^4(\sigma_0) \cos(\phi/f)$, and thus the potential
difference between two nearby minima
is estimated by $\delta V \simeq  c_1 M^4\, 2\pi f/ F \sim \mu_{\rm br}^4(v) \ll \mu_\phi^4(\sigma_0)$.
Applying the thin-wall approximation~\cite{Coleman:1980aw}, one finds that the tunneling rate 
from $\phi_\ast$ to $\phi_\ast + 2\pi n f$ is proportional to $e^{-B}$, where $B$ is roughly given by 
$B \sim 100n  f^4 \mu^8_\phi(\sigma_0) \mu^{-12}_{\rm br}(v)$. 
For reasonable values of $\mu_\phi(\sigma_0)$ and $f$, our scenario leads to $B\gtrsim 400$,
and therefore is quite safe from the vacuum decay constraint.

\begin{figure}[t]
	\begin{center}
		\begin{tabular}{l}
		\hspace{-0.4cm}
			\includegraphics[width=0.48\textwidth]{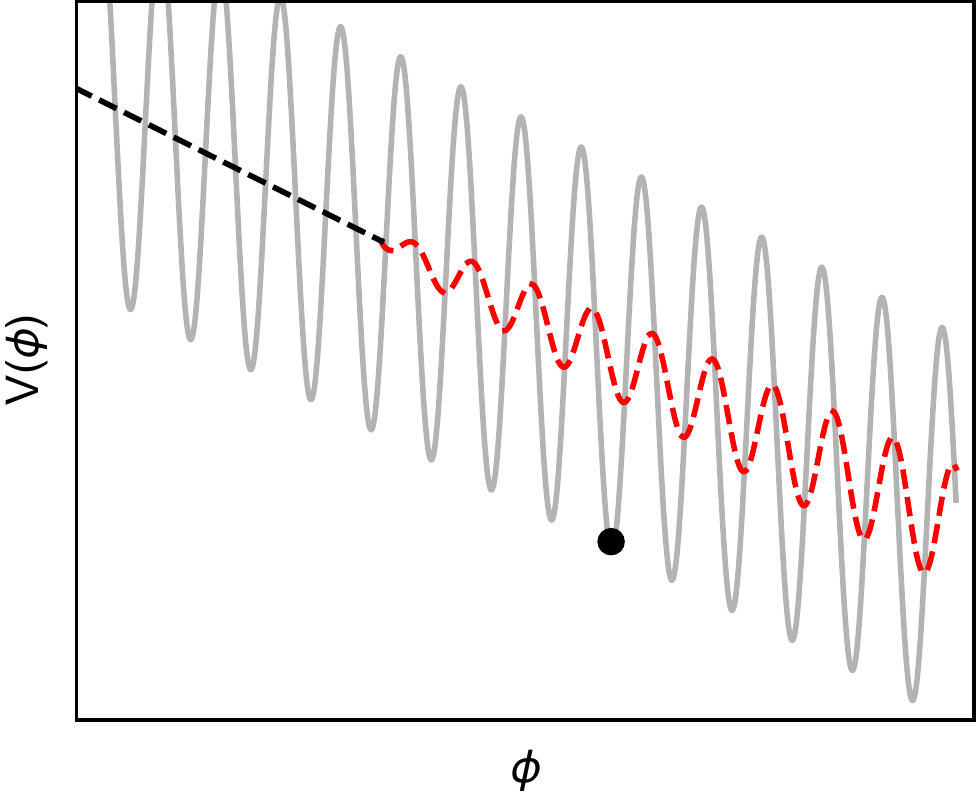}
			\hspace{0.3cm}
			\includegraphics[width=0.48\textwidth]{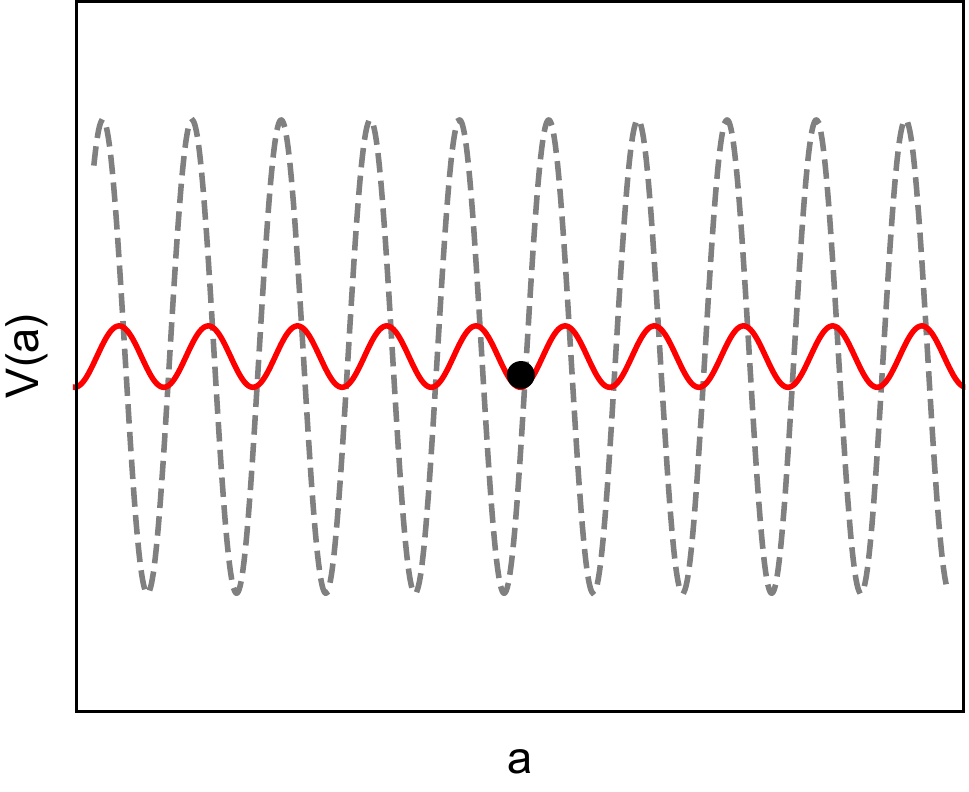}
		\end{tabular}
	\end{center}
	\caption{  
   Illustration of the cosmological relaxation of the electroweak scale in the PQ relaxion scenario where
   the QCD anomaly is responsible for selecting both the Higgs mass and the strong CP phase.
   The left (right) pannel shows the scalar potential along the direction of the relaxion  
   (the QCD axion), where the dotted (solid) curves are the potential during (after) inflation. 
   The red dotted and solid curve are the potential induced by the QCD anomaly during and 
   after inflation, respectively, while the gray curves are those generated by an inflaton-dependent 
   coupling to hidden gauge anomalies. 
   In the present universe, the relaxion (the QCD axion) is stabilized at the black dot in the left (right) pannel.
  }
	\label{fig:pots}
\end{figure}

An important issue in the relaxation mechanism is to find a viable inflation sector accommodating
very low scale inflation with a huge number of $e$-folds, which still 
requires significant progress to be made.
Low scale inflation can reproduce the observed amplitude of primordial curvature power spectrum 
for the CMB scales if there is a large hierarchy of the slow-roll parameters,
$\epsilon = M_{Pl}^2(\partial_\sigma V/V)^2 /2 \sim 10^{-30}(H_i/{\rm GeV})^2$
and $\eta = M_{Pl}^2(\partial_\sigma^2 V/V) \sim -0.01$.
This leads to consider a hybrid-type inflation with 
\bea
V(\sigma, \chi) = V_{\rm inf}(\sigma) + ( c\sigma^2 - m^2_\chi) \chi^2 + \lambda \chi^4
+ \cdots,
\eea
where inflation driven by $V_{\rm inf}$ ends when the waterfall field $\chi$ becomes tachyonic. 
In order to implement the relaxation mechanism, one can consider  
a Coleman-Weinberg potential, $V_{\rm inf} = V_0 + g^2 \xi^2 \ln (\sigma^2/M^2) + \cdots$
from the Fayet-Iliopoulos $D$-term $\xi$ with an extremely small gauge coupling $g$, 
which is realized in a supersymmetric theory with no-scale structure~\cite{Evans:2017bjs}.
However, in this case, it is nontrivial to suppress other supersymmetry breaking contributions.
Another possibility is to consider a scalar potential, $V_{\rm inf} = V_0 + m^2 \sigma^2 + \cdots$, for 
which $\eta$ is positive and thus one needs to introduce a curvaton field to generate the observed power 
spectrum~\cite{Graham:2015cka}. 
Here $m^2$ should be smaller than $H^2_i$, and it may be the consequence of 
additional (shift) symmetries. 
Note that, in the case that the field excursion of $\sigma$ is sub-Planckian, a large enough 
$e$-folds can be achieved without fine-tuning of initial conditions if $V_{\rm inf}$ is very flat.
There are interesting approaches to obtain a periodic scalar potential with a very flat plateau and sharp 
edges as a variation of natural inflation~\cite{Higaki:2015kta,Nomura:2017ehb,Gong:2017krw}. 
Such a nontrivial scalar potential is generated by extra-dimensional dynamics or a large $N$ gauge 
symmetry.  
On the other hand, our scheme is based on inflaton-dependent dynamics of the QCD axion 
and the relaxion.
If the hidden sector with the potential (\ref{inflaton-dependent-dynamics})
is responsible for $\Delta V_{\rm br}$, the inflaton mass-squared receives corrections
\bea
\delta m^2 \sim
 \left( \kappa_a + \frac{\kappa_\phi}{16\pi^2} \right) M^2,
\eea
during inflation.
Here we have used that there are quantum corrections arising from closed scalar loops,
and that $S_a$ is tachyonic during inflation and fixed at a value depending on $\sigma$.
The slow-roll conditions require $m^2 + \delta m^2 \ll \frac{M_{Pl}}{\sigma_{\rm inf}}H^2_i $. 
Combining this with the constraint (\ref{constraint-kappa}), one finds that  the relaxation can
work in the hybrid inflation setup for $\kappa_i$ lying in the range,
$(\frac{M}{\sigma_{\rm inf}})^2 < \kappa_i \ll  \frac{M_{Pl}}{\sigma_{\rm inf}} (\frac{H_i}{M})^2$,
barring fine-tunings.
It is interesting to notify that $S_\phi$ can play the role of the waterfall field.
We should also comment that some evolving scalar field other than the inflaton can 
provide the desired properties to $S_a$ and $S_\phi$.
For instance, noting that inflation is over due to a waterfall phase transition driven by $\chi$ in hybrid
inflation, one can consider 
$V = (-M^2 + \kappa^\prime_a \chi^2) |S_a|^2 + (M^2 - \kappa^\prime_\phi \chi^2)|S_\phi|^2+\cdots$ 
instead of the inflaton-dependent mass-squared terms in the potential 
(\ref{inflaton-dependent-dynamics}). 
Here $\kappa^\prime_i$ are positive constants larger than $\lambda (M/m_\chi)^2$.

We close this section by summarizing how the relaxation is implemented in our scenario.
The cosmological evolution of the relaxion $\phi$ due to the sliding potential $V_0$ changes the effective
Higgs mass-squared term from $m^2_h \sim M^2$ to a negative value.
The QCD anomaly generates Higgs-dependent barriers for $\phi$ during inflation which stop the relaxion
at the position giving $m^2_h \sim -v^2$ with $v\ll M$.
However it is a hidden confining gauge group that generates higher barriers for $\phi$ after inflation 
ends, implying that the correct selection of the effective Higgs mass-squared is not spoiled and the QCD axion 
obtains a mass from the QCD anomaly.
The cutoff scale can then be as high as about $10^7$~GeV.
On the other hand, the QCD axion acquires a large mass from a hidden strong dynamics
whose effects are turned off after inflation,
and it explains why the strong CP phase is so tiny in the present universe.
Our scheme works owing to the barrier potential $\Delta V_{\rm br}$, whose role changes dramatically
during and after inflation due to the roll of the inflaton.
Fig.~\ref{fig:pots} illustrates how the relaxation works in our scheme where the QCD anomaly is
responsible for fixing both the Higgs mass and the strong CP phase. 
Finally we note that, as in other relaxion models, the cosmological relaxation of electroweak scale 
requires a huge excursion $F\gg f$ and a long period of inflation.

\section{Stronger SM Couplings during Inflation} 

In this section, we discuss the possibility to alleviate the constraints on the relaxion excursion
and the number of $e$-folds in the relaxation mechanism. 
The idea is to make the SM couplings stronger during inflation so that the QCD effects can be enhanced,
\bea
\mu^4_{\rm br}(\sigma=\sigma_{\rm inf}) &\gg& \mu^4_{\rm br}(\sigma=\sigma_0),
\eea
where $\mu_{\rm br}(\sigma=\sigma_0,h=v) = (m_u \Lambda_{\rm QCD}^3)^{1/4}\simeq 0.1$~GeV.
Let us consider the case that the strong coupling constant $\alpha_s$ and the Yukawa couplings 
$y_i$ depend on the inflation field: 
\bea \label{eq:delta_c}
\alpha_s(\sigma_{\rm inf},M) &=& \alpha_s(\sigma_0,M) + \Delta \alpha_s,
\nonumber \\
y_i(\sigma_{\rm inf},M) &=& y_i(\sigma_0,M) + \Delta y, 
\eea  
where the inflaton-dependent contributions $\Delta \alpha_s$ and $\Delta y$ are assumed to
be positive and disppear after inflation. 
To get $\Delta \alpha_s >0$, one may introduce for instance extra quarks whose masses receive 
an additional contribution depending on the inflation field.  
The QCD then becomes strong at a higher energy scale during inflation than in the present universe.
Note that some quarks should be lighter than the QCD scale in order for the QCD anomaly to generate 
Higgs-dependent barriers for $\phi$.

Stronger SM couplings during inflation can relax the constraints on $F$ and $N_e$, but 
require the cutoff scale to be low since otherwise the relaxation mechanism would be spoiled.
The effective Higgs mass-squared term selected by the relaxion evolution would generally change 
after inflation due to $\Delta \alpha_s$ and $\Delta y$.
The variation can be estimated using its sensitivity to the cutoff scale
\bea
m^2_h = \left(
6 \lambda_h -6\sum_i y^2_i  + \frac{3}{4} g^2_Y + \frac{9}{4} g^2_2 +\cdots
\right) \frac{M^2}{16\pi^2},
\eea
for the Higgs quartic coupling $\lambda_h$ and SM gauge couplings $g_Y$ and $g_2$. 
Here the ellipsis indicates terms from higher loops. 
For the given change of the couplings in Eq.~(\ref{eq:delta_c}), 
the relaxation works if 
\bea
|\Delta m^2_h| \simeq \left| 
\frac{\Delta y }{y_t} -  0.1\Delta \alpha_s \right| \frac{3 y_t^2 M^2}{4\pi^2} \lesssim \, v^2,
\eea
where we have taken $\Delta y \ll y_t$ with $y_t$ being the top quark Yukawa coupling,
and included  two-loop quadratic divergent contributions to the Higgs mass 
squared~\cite{Alsarhi:1991ji,Hamada:2012bp}.
The above relation shows that the cutoff scale should be
\bea
M \lesssim  
\frac{10\,{\rm TeV}}{\sqrt{|\Delta y |/0.01 + |\Delta\alpha_s|/0.1}},\quad
\eea 
barring accidental cancellation. 
The constraint on $M$ is stronger than that required for the relaxation unless $\Delta \alpha_s$
and  $\Delta y$ are very tiny.
In the presence of such inflaton-dependent contributions, the QCD-induced barriers can be
enhanced by 
\bea
\mu_{\rm br}(\sigma_{\rm inf},v)
\simeq
10\,{\rm GeV}
\left(\frac{y_u + \Delta y }{0.01}\right)^{1/4}
\left(\frac{\Lambda_{\rm QCD}(\sigma_{\rm inf})}{20\,{\rm GeV}}\right)^{3/4},
\eea  
in which the QCD scale during inflation is given by
\bea 
\Lambda_{\rm QCD}(\sigma_{\rm inf}) = 
 {\rm Exp}\left[\frac{2\pi}{11 -  2n_f/3}\left(\frac{1}{\alpha_s(\sigma_0, M)} 
 - \frac{1}{\alpha_s(\sigma_0, M)+ \Delta \alpha_s}\right)\right]\Lambda_{\rm QCD},
\eea
where $n_f$ counts the number of SM quarks lighter than $\Lambda_{\rm QCD}(\sigma_{\rm inf})$.
For instance, one obtains  $\Lambda_{\rm QCD}(\sigma_{\rm inf})\simeq 60\,\Lambda_{\rm QCD}$
for $\Delta \alpha_s=0.2$ and $n_f= 3$.
Note that the QCD anomaly leads to $\mu^4_{\rm br} \propto h$ if $n_f\neq 0$, that is
for $\Delta y \lesssim 2\times 10^{-3}\, \Lambda_{\rm QCD}(\sigma_{\rm inf})/\Lambda_{\rm QCD}$.

Let us examine the requirements for the relaxation mechanism.
The inflation scale should be constrained to be
\bea
H_i < 0.2\,{\rm GeV}  
\left(\frac{f}{10^6\,{\rm GeV}}\right)^{-1/3}
\left(\frac{\mu_{\rm br}(\sigma_{\rm inf},v) }{10\,{\rm GeV}}\right)^{4/3}.
\eea
In addition, one needs $F$ much larger than $f$:
\bea
\frac{F}{f} \sim 0.6\times 10^{10} 
\left( \frac{c_1}{1/16\pi^2} \right)
\left( \frac{\mu_{\rm br}(\sigma_{\rm inf},v) }{10\,{\rm GeV}} \right)^{-4}
\left( \frac{M}{10\, {\rm TeV}} \right)^4,
\eea
showing that the required hierarchy between $F$ and $f$ is reduced due to stronger
SM couplings during inflation.
It also follows that the relaxation can be implemented without a huge number of
$e$-folds:  
\bea
N_e > \left(\frac{F}{M_{Pl}}\right)^2
\sim
20 \left(\frac{F/f}{10^{13}} \right)^2\left(\frac{f}{10^6\,{\rm GeV}}\right)^2,
\eea
for the cutoff scale $M$ lower than about $10$~TeV.

\section{Conclusions}

In this paper, we have explored the Peccei-Quinn relaxion scenario, in which the QCD anomaly is responsible
for selecting the Higgs mass as well as the strong CP phase.  
Our scenario involves two axion-like scalars, the relaxion and the QCD axion, which couple simultaneously
to the QCD anomaly, and also separately to a hidden gauge anomaly via inflaton-dependent dynamics.
During inflation, the QCD axion becomes heavy due to the hidden strong force, and
the Higgs-dependent scalar potential induced by the QCD anomaly plays the role of a back-reaction potential 
implementing the relaxation of the electroweak scale. 
On the other hand, after inflation, the hidden strong force makes the relaxion heavy so that
the QCD axion can be stabilized dominantly by the QCD-induced potential, yielding a vanishing
strong CP phase.  
In our scheme, the relaxation mechanism is compatible with reheating temperature higher than 
the electroweak scale, 
and the QCD axion becomes a natural candidate for dark matter, whose amount is mostly determined 
by the misalignment mechanism.  
We have also studied a more general case where the Yukawa and gauge couplings of the SM 
are dependent on the field value of the inflaton.  
If the SM couplings become stronger during inflation, the relaxation can be achieved by
a sub-Plankian field excursion of the relaxion  
during inflation for a cutoff scale below $10$~TeV. 
This would imply that the relaxion may play a partial role in stabilizing the electroweak scale
against radiative corrections and need helps from other physical effects such as supersymmetry.

\acknowledgments

KSJ is supported by the National Research Foundation of Korea (NRF) grant funded by the Korea 
government (MSIP) (NRF-2015R1D1A3A01019746).
CSS acknowledges  the support from the Korea Ministry of Education, Science and Technology, 
Gyeongsangbuk-Do and Pohang City for Independent Junior Research Groups at the Asia Pacific Center 
for Theoretical Physics. 
CSS was also supported by the Basic Science Research Program through the NRF Grant 
(No.~2017R1D1A1B04032316). 
CSS is supported by IBS under the project code, IBS-R018-D1.

\bibliographystyle{JHEP}

\bibliography{pqr}

\end{document}